\documentclass[11pt]{article}
\usepackage[utf8]{inputenc}
\usepackage{fullpage}
\usepackage{authblk}
\usepackage{natbib}
\RequirePackage[colorlinks=true,allcolors=blue,hypertexnames=false]{hyperref}%
\usepackage{graphicx,todonotes}
\usepackage{outlines}
\usepackage{array}
\usepackage{selectp}
\usepackage{float}

\usepackage{courier}

\usepackage{titlesec}
\AtBeginDocument{%
  \providecommand\BibTeX{{%
    \normalfont B\kern-0.5em{\scshape i\kern-0.25em b}\kern-0.8em\TeX}}}

\usepackage{./Latex/macros_01_proof_envs}
\usepackage{./Latex/macros_02_operators_commands}
\usepackage{./Latex/macros_03_symbols}
\usepackage{./Latex/macros_04_propernames}
\usepackage{./Latex/macros_05_reviewer}

\DeclareRobustCommand{\rchi}{{\mathpalette\irchi\relax}}
\newcommand{\irchi}[2]{\raisebox{\depth}{$#1\chi$}}

\usepackage{listings}
\usepackage{color} 
\definecolor{dkgreen}{rgb}{0,0.6,0}
\definecolor{gray}{rgb}{0.5,0.5,0.5}
\definecolor{mauve}{rgb}{0.58,0,0.82}
\newcommand\blfootnote[1]{%
  \begingroup
  \renewcommand\thefootnote{}\footnote{#1}%
  \addtocounter{footnote}{-1}%
  \endgroup
}
\usepackage{enumitem}

\usepackage{listingsutf8}

\crefname{lstlisting}{\listingname}{\listingname s}
\Crefname{lstlisting}{\listingname}{\listingname s}

\usepackage{verbatim}

\title{\bf \texttt{maars:} Tidy Inference under the `Models as Approximations' Framework in \texttt{R}}
\author{Riccardo Fogliato$^*$}
\author{Shamindra Shrotriya$^*$}
\author{Arun Kumar Kuchibhotla}
\affil{Department of Statistics \& Data Science, Carnegie Mellon University}
\affil{\texttt{\{rfogliat,sshrotri,akuchibh\}@andrew.cmu.edu}}

\date{}                   
\begin{document}
\maketitle

\begin{abstract}
  Linear regression using ordinary least squares (OLS) is a critical part of every
    statistician's toolkit. In R, this is elegantly implemented via \texttt{lm()} and its
    related functions. However, the statistical inference output from this suite of
    functions is based on the assumption that the model is well specified. This
    assumption is often unrealistic and at best satisfied approximately. In the
    statistics and econometrics literature, this has long been recognized and a large
    body of work provides inference for OLS under more practical assumptions. This can
    be seen as model-free inference. In this paper, we introduce
    our package \texttt{maars} (``models as approximations'') that aims at bringing
    research on model-free inference to \texttt{R} via a comprehensive workflow.
    The \texttt{maars} package differs from other packages that also implement
    variance estimation, such as \texttt{sandwich}, in three key ways. First, all functions
    in \texttt{maars} follow a consistent grammar and return output in tidy format, with minimal deviation from the typical
    \texttt{lm()} workflow. Second, \texttt{maars} contains several tools for
    inference including empirical, multiplier, residual bootstrap, and subsampling, for easy comparison. Third,
    \texttt{maars} is developed with pedagogy in mind. For this, most of its
    functions explicitly return the assumptions under which the output is valid.
    This key innovation makes \texttt{maars} useful in teaching inference under
    misspecification and also a powerful tool for applied researchers. We hope our
    default feature of explicitly presenting assumptions will become a de facto
    standard for most statistical modeling in \texttt{R}.\blfootnote{$^*$ Indicates equal
    contribution.}
\end{abstract}

\section{Introduction}\label{sec:introduction}
Multiple linear regression is one of the most commonly used data analytic tools.
Although the general notion of ``all models are approximations'' is well
received in terms of inference under model misspecification, most statistical
software (including \texttt{R}) does not yet have a comprehensive implementation
of all such aspects. In detail, the elegant implementation of the ordinary least squares
estimator through the \texttt{lm()} function is based on the classical and
unrealistic assumptions of a linear model, \ie, linearity, homoscedasticity, and
normality. The \texttt{lm()} implementation does provide methods for diagnosing
these assumptions. However, it has long been recognized that inference should be
performed under more general assumptions, even if the estimator is constructed
based on likelihood
principles~\citep{huber1967behavior,buja2019modelsasapproximationspart1}. 

Over the years, several \texttt{R} packages such as \texttt{car}
\citep{car2019cran}, \texttt{clubSandwich} \citep{clubsandwich2021cran},
\texttt{lmtest} \citep{lmtest2002cran}, and
\texttt{sandwich} \citep{sandwich2004lmjss,sandwich2020jssversatilevariances} 
have been developed
to implement heteroscedasticity and non-linearity consistent estimators of the
asymptotic variance. In this misspecified setting, a \textit{tidy workflow} is currently
unavailable in \texttt{R}. By a tidy workflow, we mean a suite of
functions for variance estimation, confidence interval computation, hypothesis
testing, and diagnostics that follow a consistent naming convention, whose
outputs are tailored for further exploration, \eg, using the \texttt{tidyverse} 
set of packages \citep{tidyverse2019cranjoss}. In this paper, we introduce
the \texttt{maars} package that provides such a tidy workflow for
data analysis with ordinary least squares (OLS) linear regression. Our package
is motivated, in large part, by the discussion
of~\citet{buja2019modelsasapproximationspart1,buja2019modelsasapproximationspart2}.
Part of our motivation is also pedagogical, in making applied researchers more 
explicitly aware of the underlying assumptions and the consequences of 
model-based variance estimators for inference. 

In this paper, we introduce the key functionality in the
\texttt{maars} package.
For illustrative purposes only, we use the \textit{LA Homeless data}
from~\cite{buja2019modelsasapproximationspart1} as a running example to 
demonstrate the \texttt{maars} package functionality.\footnote{The package is available at:
\url{https://github.com/shamindras/maars}.}
As noted in~\cite{buja2019modelsasapproximationspart1}, this dataset has $505$
observations, each representing a sampled metropolitan LA County Census tract.
It also has $7$ numeric variables measuring different quantities of interest in
each tract. For linear modeling purposes, the response variable
(\textit{StreetTotal}) is a count of the homeless people in each tract. There
are six covariates for regression purposes. These include the Median Income of
Households (\textit{MedianInc (\$1000})), and the share of non-Caucasian
residents  
(\textit{PercMinority}). The remaining four covariates measure the proportion of
the different types of lots in each tract (\textit{PercCommercial},
\textit{PercVacant}, \textit{PercResidential} and \textit{PercIndustrial}).

The rest of the paper is organized as follows. In \Cref{sec:theoretical-properties-ols}, 
we detail the theoretical properties of the OLS estimator in both the well-specified
and misspecified model settings.
In \Cref{sec:guiding-design-principles}, we describe the key design principles
upon which \texttt{maars} is based.
\Cref{sec:variance-estimation-maars} and \Cref{sec:diagnostic-tools} 
focus on the variance estimation and graphical model diagnostic functionalities present in 
the package respectively.
In \Cref{sec:teaching-by-example-vignettes-grammar},
we give an overview of the pedagogical vignettes and and guided lesson-plans in the package.
In \Cref{sec:open-source-best-practices}, we discuss the open-source 
best practices that we adopted to help build an inclusive community for 
new users and contributors. Finally, in \Cref{sec:conclusion-and-future-work} 
we conclude by outlining future development plans.

\section{Theoretical properties of OLS}\label{sec:theoretical-properties-ols}
\subsection{OLS under well-specification} 
\begin{figure}[H]
    \centering
    \includegraphics[width=0.5\linewidth]{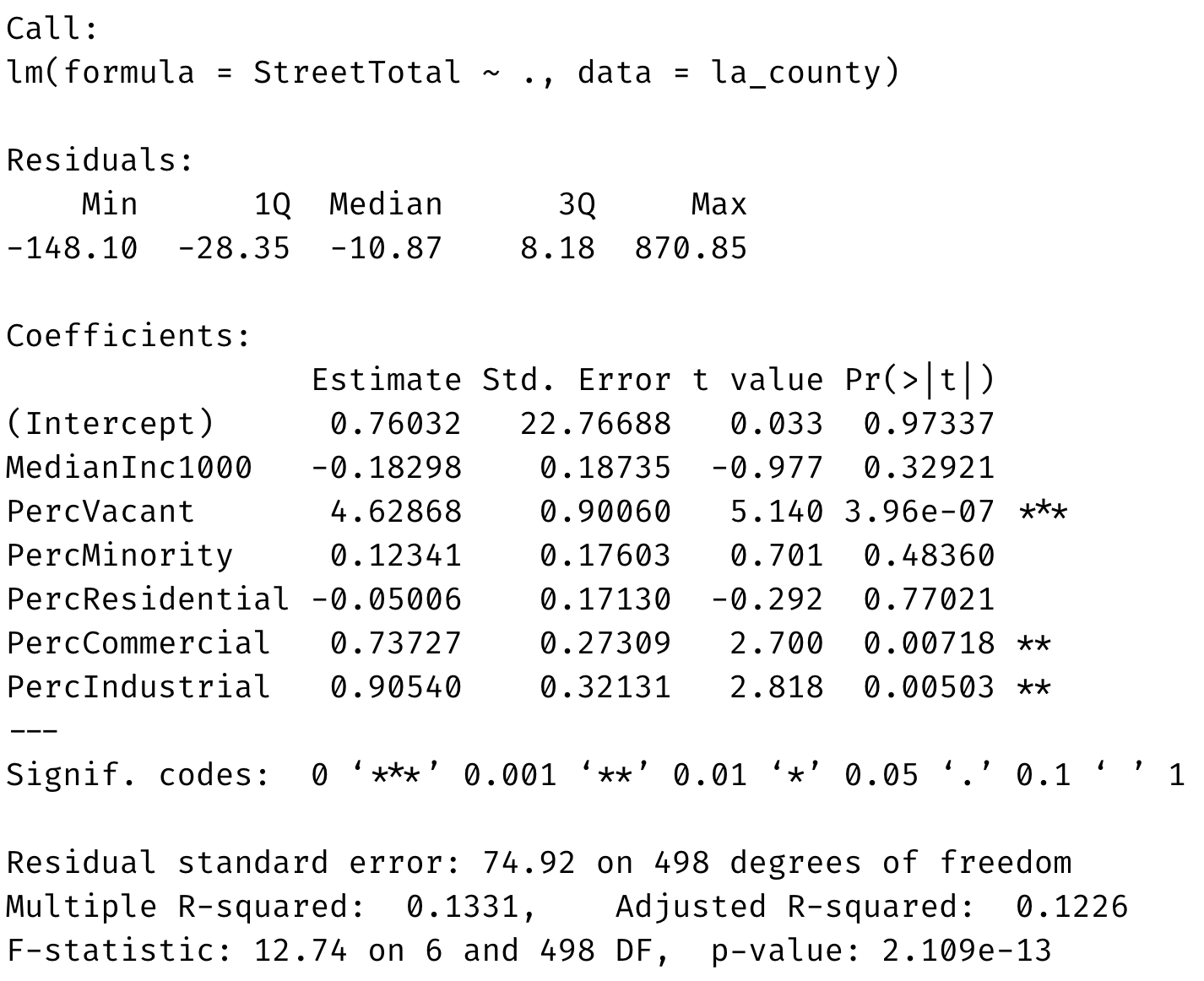}
    \caption{Output of \texttt{summary(mod\_fit)}.}
    \label{fig:mod-fit-summary-output}
\end{figure}

Suppose we have regression data $(X_i,
Y_i)\in\reals^d\times\reals, 1\le i\le n$. The well-specified (classical) linear
model stipulates that $(X_i, Y_i), 1\le i\le n$ are independent and satisfy
\begin{align}\label{eq:linear-model} Y_i &= X_i^{\top}\beta_0 + \varepsilon_i,
\end{align} where $\varepsilon_i|X_i\overset{\text{i.i.d.}}{\sim} N(0,\sigma^2)$. This implies that
$\mbbE \left[ Y_i|X_i\right]  = X_i^{\top}\beta_0$ (linearity) and $\mbox{Var}(Y_i|X_i) =
\sigma^2$ (homoscedasticity). Usually one also assumes covariates to be
non-stochastic. Under these assumptions, we get that the OLS estimator
$$\widehat{\beta} = \argmin_{\theta\in\reals^d}\sum_{i=1}^n (Y_i -
X_i^{\top}\theta)^2,$$ has a normal distribution (conditional on $X_i, i\le n$):
\begin{align}\label{eq:distribution-well-specified} \widehat{\beta} -
\beta_0 ~\sim~ \distNorm\left(0, \frac{\sigma^2}{n}\widehat{J}^{-1}\right)\quad\mbox{where}\quad \widehat{J} := \frac{1}{n}\sum_{i=1}^n X_iX_i^{\top}.
\end{align} This distribution yields confidence intervals, hypothesis tests, and
$p$-values for $\beta_0$. The \texttt{R} function \texttt{lm()} provides
inference based on~\eqref{eq:distribution-well-specified}. For the LA Homeless
data, the implementation of OLS along with its inference 
(in \Cref{fig:mod-fit-summary-output}) are as follows:

\begin{lstlisting}[language=R, title={}, caption={}, basicstyle=\footnotesize\ttfamily,
label={nlst:r00_mod_lm_fit},numbers=none]
# load LA county homeless data into dataframe 
la_county <- get(data("la_county", package = "maars"))

# fit OLS on LA county data
mod_fit <- lm(formula = StreetTotal ~ ., data = la_county)

# summary of fitted model
summary(mod_fit)
\end{lstlisting}

As mentioned, the standard errors, $t$-scores, and the $p$-values reported in
the summary above are most likely invalid because the linear model for LA Homeless
data is not well-specified~\citep{buja2019modelsasapproximationspart1}. We will
now discuss the properties of OLS under misspecification of the linear model. It
should be mentioned here that we will only discuss misspecification of the
linear model but require independence of the observations. Dependence will be
part of our future work (see \Cref{sec:conclusion-and-future-work}).
\subsection{OLS under misspecification} If $(X_i, Y_i)\in\reals^d\times\reals,
1\le i\le n$ are independent and identically distributed (\iid), then the least
squares estimator $\widehat{\beta}$ converges (in probability) to 
\begin{align}
    \beta^{\infty} ~\defined~
    \argmin_{\theta\in\reals^d}\,\Exp{(Y - X^{\top}\theta)^2}, 
\end{align} 
where $(X,
Y)\in\reals^d\times\reals$ is an independent copy of $(X_i, Y_i)$ and $\beta^{\infty}$  $(\widehat{\beta}\mbox{ at }n=\infty)$ is the population
projection parameter. If the linear
model~\eqref{eq:linear-model} holds true, then $\beta^{\infty} = \beta_0$.
Assuming only \iid observations with finite moments, it is easy to obtain the
normal limiting distribution: 
\begin{align}
    \sqrt{n}(\widehat{\beta} - \beta^{\infty})
    ~\convd~ N(0, J^{-1}VJ^{-1}),
\end{align}
where $J = \Exp{XX^{\top}}$, and $V = \Exp{XX^{\top}(Y -
X^{\top}\beta^{\infty})^2}$; see~\cite{buja2019modelsasapproximationspart1} for
details. A similar limit theorem also holds when the observations are
independent but not identically distributed (\inid). 

The matrices $J$ and $V$
can be readily estimated by replacing the expectations by averages and
$\beta^{\infty}$ by $\widehat{\beta}$ leading to the sandwich estimator
$\widehat{J}^{-1}\widehat{V}\widehat{J}^{-1}$; see \cite{kuchibhotla2018model} for details.
Here 
\begin{align}\label{eq:Jhat-Vhat} 
    \widehat{J} = \frac{1}{n}\sum_{i=1}^n
    X_iX_i^{\top}\quad\mbox{and}\quad \widehat{V} = \frac{1}{n}\sum_{i=1}^n
    X_iX_i^{\top}(Y_i - X_i^{\top}\widehat{\beta})^2. 
\end{align} It can be shown
that $\widehat{J}^{-1}\widehat{V}\widehat{J}^{-1}$ consistently estimates the
asymptotic variance under \iid data and over-estimates the asymptotic variance under
\inid data \citep{kuchibhotla2018model}. It is noteworthy that the estimator
$\widehat{J}^{-1}\widehat{V}\widehat{J}^{-1}$ converges to the model-based
variance if the linear model~\eqref{eq:linear-model} holds true.
With a model-free estimator of the asymptotic variance, one can perform inference for $\beta^\infty$ using asymptotic normality. 
The asymptotic $(1-\alpha)$ confidence interval for the $j^{th}$ coefficient $\beta^\infty(j)$ is then given by
\begin{align}
    \widehat{\mathrm{CI}}_j = \left[\widehat{\beta}(j)-z_{\alpha/2}\sqrt{\frac{(\widehat{J}^{-1}\widehat{V}\widehat{J}^{-1})_{jj}}{n}}, \widehat{\beta}(j)+z_{\alpha/2}\sqrt{\frac{(\widehat{J}^{-1}\widehat{V}\widehat{J}^{-1})_{jj}}{n}}\right],
\end{align}
where $z_{\alpha/2}$ is the quantile of the standard normal.
The Wald test for $k$ linear restrictions of the coefficients of the form $R\beta=r$, where $R$ and $r$ are a $k\times d$ matrix and a $k$ dimensional vector respectively, is still valid under only the assumption of independence. 
Under the null hypothesis, the Wald test statistic $(R\widehat\beta-r)^T (R\widehat{J}^{-1}\widehat{V}\widehat{J}^{-1}R^T)^{-1}(R\widehat\beta-r)$ asymptotically follows a $\rchi^2$ distribution with $k$ degrees of freedom; see, \eg, \citet[Theorem~3.4]{white1982maximum}.

\section{\texttt{maars}: Designed for research and pedagogy}\label{sec:guiding-design-principles}

Our goal is for the \texttt{maars} package to be a natural research workflow for
model-free OLS inference in \texttt{R}. Equally importantly, we also intend for
\texttt{maars} to be a useful pedagogical tool to help convey these rich
inferential concepts. In order to satisfy this dual research and pedagogical
purpose, \texttt{maars} is built according to the following five design
principles (\ref{itm:dp1}--\ref{itm:dp5}):

\benum[label=\textbf{DP.\arabic*}, align=left, start=1]
    \item \label{itm:dp1} Build a flexible API to enable easy user access to a
    diverse set of inferential tools. 
    \item \label{itm:dp2} Ensure that the assumptions for all modeling techniques 
    explicit to the user.
    \item \label{itm:dp3} Develop a consistent \emph{tidy grammar} for 
    OLS inference under model misspecification to enhance communication 
    of code with other users and developers. 
    \item \label{itm:dp4} Emphasize teaching these inferential techniques and 
    concepts through instructive examples. 
    \item \label{itm:dp5} Utilize modern open-source best practices to create an
    inclusive package development environment.
\eenum

We provide brief details on the design principles~\ref{itm:dp1}--\ref{itm:dp5}
and details of how they are applied in \texttt{maars}. First, by~\ref{itm:dp1}, 
we intend for \texttt{maars} to provide a 
user-friendly way to access the rich set of OLS inferential techniques under 
model misspecification. This is primarily achieved through the \texttt{comp\_var()} 
function, which provides a common API for the user for OLS modeling purposes. 
By~\ref{itm:dp2} we want to ensure that the assumptions behind the modeling 
techniques are made explicit to the user. By~\ref{itm:dp3} we have adopted in 
\texttt{maars} a consistent function naming convention that follows a tidy grammar 
and where tidy tibble outputs are returned whenever feasible. This is to better 
facilitate downstream analysis and communication of results. Details 
of~\ref{itm:dp1}--\ref{itm:dp3} are emphasized with examples in \Cref{sec:variance-estimation-maars,sec:diagnostic-tools}.

In~\ref{itm:dp4} we emphasize the use of detailed instructive vignettes to teach
the inferential techniques for OLS using a tidy \texttt{maars} workflow. See~\Cref{sec:teaching-by-example-vignettes-grammar} for more details.
Finally, in~\ref{itm:dp5} our aim is for \texttt{maars} to be a long-term part
of a applied researcher's OLS toolkit in \texttt{R}. We achieve this by following
and adapting to the best open-source package development practices, and thus
building an inclusive environment around \texttt{maars}. For example, we ensure
that the \texttt{maars} functions are rigorously benchmarked and unit tested. A
well defined benchmarking framework demonstrates how our code is optimized for
computational efficiency. A detailed unit testing framework allows us to
effectively perform statistical validation and error handling. See
\Cref{sec:open-source-best-practices} for details. Finally, we note that the
\texttt{maars} package is developed in an open-source manner that is inclusive
towards new users and contributors.

\section{Variance estimation and inference in \texttt{maars}}\label{sec:variance-estimation-maars}

Under model misspecification, one of the main inferential tools is accurate and efficient estimation of
variance. The \texttt{maars} modeling framework offers a variety of OLS
inferential tools that are readily usable as part of a tidy pipe-friendly
(\texttt{\%>\%}) workflow \citep{milton2020magrittrcran}. 

These tools fall into three categories.
First, we have closed-form variance estimators that are computed by default,
\ie, \texttt{lm()} and the sandwich standard errors. Second, we have
resampling-based variance estimators. These include the empirical bootstrap,
multiplier bootstrap, residual bootstrap, and the subsampling-based methods.
Third, we provide by default valid hypothesis testing under OLS model
misspecification via $\rchi^{2}$ tests. The workhorse that enables the simultaneous
computations of the several types of variance estimates is the
\texttt{comp\_var()} function, which provides a common API for the user to
access these estimators. For the sake of brevity, 
in \Cref{sec:closed-form-variance-estimation} and \Cref{sec:resampling-based-variance} we 
show how the variance estimators can be run by the user, without
displaying individual code outputs. In 
\Cref{subsec:putting-it-all-together-summary-results} we present how the consolidated
output summary of all computed variance estimators can be used for downstream inference. We now review the functionality behind each 
estimation method in turn in this section, starting with the 
closed-form variance estimators. 

\subsection{Closed-form variance estimation}\label{sec:closed-form-variance-estimation}
In \texttt{maars}, we provide two closed form variance estimators: one based on the assumption of a well-specified linear model and another based on the asymptotic normality under potential misspecification of the linear model. 

Under the ``models as approximations'' framework, using only the independence of observations
and finite moment assumptions, the asymptotic variance of the OLS estimates can be estimated by the sandwich estimator~\citep{buja2019modelsasapproximationspart1, buja2019modelsasapproximationspart2}:
    $\widehat{\mathrm{Var}}(\widehat{\beta})= n^{-1}\widehat{J}^{-1}\widehat{V}\widehat{J}^{-1},$
with $\widehat{J}$ and $\widehat{V}$ defined in~\eqref{eq:Jhat-Vhat}.
Then the standard error for the $j^{th}$ coefficient is given by 
    $\widehat{\mathrm{SE}}_j(\widehat{\beta})= n^{-1/2}(\widehat{J}^{-1}\widehat{V}\widehat{J}^{-1})_{jj}^{{1}/{2}}$.
The classical estimator of asymptotic variance under a well-specified linear model is $n^{-1}\widehat{\sigma}^2\widehat{J}^{-1}$, for the residual sum of squares based estimator $\widehat{\sigma}^2$. 
In the \texttt{maars} package, this and other inferential summary
statistics can be readily computed by calling the \texttt{comp\_var()} function as follows.

\begin{lstlisting}[language=R, title={}, caption={}, basicstyle=\footnotesize\ttfamily,
label={nlst:r01-se-sand},numbers=none]
# obtain maars with obj. w/ sandwich variance
maars_var_sand <- mod_fit %>% comp_var()
\end{lstlisting}

\nit The sandwich estimator is always computed by default, even when no arguments are passed 
in \texttt{comp\_var()}. In the above code, \texttt{maars\_var\_sand} is
an object of class \texttt{maars\_lm}. This new object represents an modified version of 
the \texttt{mod\_fit} object of class \texttt{lm} augmented with the newly computed 
sandwich variance estimates. As a result, \texttt{comp\_var()} also returns by default the well-specified OLS standard error output.

\subsection{Resampling-based variance estimation}\label{sec:resampling-based-variance}
In this subsection, we discuss the resampling-based variance estimation tools in \texttt{maars}. 
These resampling-based variance estimators are not computed by default
in \texttt{comp\_var()} and instead require a list of parameter inputs to be
passed in by the user. We now describe the details each of the methods in turn.

\subsubsection{Empirical bootstrap}\label{subsec:empirical-bootstrap-variance}
Under the independence and finite moment assumptions, another method for the
estimation of variance is the empirical bootstrap. The $m$-out-of-$n$ empirical
bootstrap consists of two steps. In the first step,
for $1\le b\le B$, we resample with replacement $m$ observations from the
original data set and obtain the OLS estimate $\widehat{\beta}_b^*$ on the
bootstrapped data set.  
In the second step, we compute $B$ times the sample covariance of
$\widehat{\beta}^*_b, 1\le b\le B$ as an estimate of ${J}^{-1}{V}{J}^{-1}$, the
asymptotic variance of $\sqrt{n}(\widehat{\beta} - \beta^{\infty})$. The
justification for this method follows from the fact that, conditional on the
original data $\{(X_i,Y_i)\}_{i=1}^n$, we have
\begin{align}
    \sqrt{m}(\widehat{\beta}_b^*-\widehat{\beta})~\convd~ \distNorm(0,\widehat{J}^{-1}\widehat{V}\widehat{J}^{-1}) \text{ as }m\rightarrow\infty
\end{align}
and thus we can show that
\begin{align}
    \frac{m}{B-1}\sum_{b=1}^B(\widehat{\beta}_b^*-\overline{\widehat{\beta}^*}) (\widehat{\beta}_b^*-\overline{\widehat{\beta}^*})^{T}~\xrightarrow{p}~ \widehat{J}^{-1}\widehat{V}\widehat{J}^{-1} \text{ as }B\rightarrow\infty.
\end{align}
Here $\overline{\widehat{\beta}^*} = B^{-1}\sum_{b=1}^B \widehat{\beta}^*_b$.
Therefore, as $B$ increases, the covariance of the OLS estimates on the bootstrapped data sets converges 
in probability to $\widehat{J}^{-1}\widehat{V}\widehat{J}^{-1}$. We refer the reader to~\cite{bickel1997resampling} for more details.
In \texttt{maars}, the $n$-out-of-$n$ empirical bootstrap estimates for the 
LA Homeless data can be obtained as follows. In this dataset, there are $n = 505$
observations and hence setting $m = 505$ is the same as running the $n$-out-of-$n$
empirical bootstrap.

\begin{lstlisting}[language=R, title={}, caption={}, basicstyle=\footnotesize\ttfamily,
label={nlst:r02-se-emp},numbers=none]
# obtain maars obj. w/ emp. boot. variance
maars_var_emp <- mod_fit %>%
  comp_var(boot_emp = list(B = 1e3, m = 505))
\end{lstlisting}

\subsubsection{Multiplier bootstrap}\label{subsec:multiplier-bootstrap-variance}

The main disadvantage of the empirical bootstrap is that it requires the 
computation of the OLS estimate $B$ times. Furthermore, the bootstrapped design 
matrix might be singular. 
The multiplier bootstrap is an alternative method that helps overcome these difficulties. The method consists 
of estimating $B$ times
\begin{align}\label{eq:multiplier-bootstrap}
    \widehat{\beta}^*_b=\widehat{\beta}+\frac{1}{n}\sum_{i=1}^n w_{i,b} \widehat{J}^{-1}X_i (Y_i-X_i^{\top}\widehat{\beta}),\quad 1\le b\le B.
\end{align}
The weights $\{w_{i,b}\}_{i=1}^{n}$ are independent with $\Exp{w_{i,b}}=0$ and $\Var{w_{i,b}}=1$,
for each $1\le i \le n$, $1\le b \le B$. 
These can be computed as follows:
\begin{lstlisting}[language=R, title={}, caption={}, basicstyle=\footnotesize\ttfamily,
label={nlst:r03-se-mult},numbers=none]
# obtain maars obj. w/ mul. boot. variance
maars_var_mul <- mod_fit %>% 
	comp_var(boot_mul = list(B = 1e3, weights_type = "rademacher"))
\end{lstlisting}

The distribution of the weights in~\eqref{eq:multiplier-bootstrap} is specified through the \texttt{weights\_type} parameter. Following the \texttt{boottest} package~\citep{roodman2019fastwildinferencestataboottest} in \texttt{Stata},
we provide the user with
four prespecified options: 
\texttt{rademacher}, \texttt{mammen}, 
\texttt{webb}, \texttt{gaussian}.
The \texttt{rademacher} weights are sampled from the two 
point Rademacher distribution which has takes values $\theseta{-1, 1}$,
with equal probability. The \texttt{mammen} weights are sampled from 
the two point Mammen distribution which takes values 
$(\mp\sqrt{5} + 1)/2$, with probabilities 
$(\sqrt{5}\pm1)/(2\sqrt{5})$, respectively. 
The \texttt{webb} weights are sampled from the 
six point Webb distribution which takes values 
$\{\pm \sqrt{{3}/{2}}, \pm \sqrt{{1}/{2}},\pm 1\}$, 
with equal probability. Finally, the \texttt{gaussian} weights 
are sampled from the standard Gaussian distribution.

\subsubsection{Residual bootstrap}
One of the classical resampling methods for linear regression is residual bootstrap. In this method, one resamples the residuals instead of the pairs $(X_i, Y_i)$ as in empirical bootstrap. With $\widehat{\beta}$ representing the OLS estimator, let the residuals be denoted by $e_i = Y_i - X_i^{\top}\widehat{\beta}$. Residual bootstrap consists of estimating $B$ times
\[
\widehat{\beta}_b^* ~:=~ \argmin_{\theta\in\mathbb{R}^d}\,\frac{1}{n}\sum_{i=1}^n (Y_{i,b}^* - X_i^{\top}\theta)^2,
\]
where $Y_{i,b}^* = X_i^{\top}\widehat{\beta} + e_{i,b}^*, 1\le i\le n$ for $e_{i,b}^*, 1 \le i\le n$ are drawn i.i.d. from the empirical distribution of $e_1, \ldots, e_n$. From the closed form expression of $\widehat{\beta}_b^*$, it readily follows that conditionally on the original data $(X_i, Y_i), 1\le i\le n$, as $n\to\infty$,
$\sqrt{n}(\widehat{\beta}^*_b - \widehat{\beta}) \overset{d}{\to} N(0, \widehat{\sigma}^2\widehat{J}^{-1}),$
where $\widehat{\sigma}^2 = n^{-1}\sum_{i=1}^n e_i^2$. This convergence in distribution does not require any of the linear model assumptions.
However, the asymptotic variance of the residual bootstrap estimator matches the linear model based closed-form variance estimator and is not valid under model misspecification. In \texttt{maars}, residual bootstrap variance estimator can be obtained as follows: 


\begin{lstlisting}[language=R, title={}, caption={}, basicstyle=\footnotesize\ttfamily,
label={nlst:r03-se-res},numbers=none]
# obtain maars obj. w/ res. boot. variance
maars_var_res <- mod_fit %>%
  comp_var(boot_res = list(B = 1e3))
\end{lstlisting}

\subsubsection{Subsampling} 
An alternative to bootstrap is subsampling~\citep{politis1994large}. Given our
discussion on empirical ($m$-out-of-$n$) bootstrap, subsampling can be described
as the $m$-out-of-$n$ bootstrap where samples are drawn \textit{without}
replacement from the original data. Recall that the $m$-out-of-$n$ bootstrap draws
samples with replacement. Unlike the $m$-out-of-$n$ bootstrap where $m$ can be of
the same order as $n$ or even larger, subsampling validity usually requires $m =
o(n)$. We refer the reader to~\cite{politis1994large} for details. Subsampling
is more generally applicable than empirical and multiplier bootstrap schemes,
allowing for valid inference even with stationary dependent data.
In \texttt{maars}, subsampling can be run as follows:

\begin{lstlisting}[language=R, title={}, caption={}, basicstyle=\footnotesize\ttfamily,
label={nlst:r03-se-sub},numbers=none]
# obtain maars obj. w/ subsampling variance
maars_var_sub <- mod_fit %>%
  comp_var(boot_sub = list(B = 1e3, m = 200))
\end{lstlisting}

\subsection{Putting it all together: Summarizing the results}\label{subsec:putting-it-all-together-summary-results}

In \Cref{sec:closed-form-variance-estimation} and \Cref{sec:resampling-based-variance}, we have
demonstrated how the \texttt{maars} package can be used to compute the variance
of the coefficients estimates based on the sandwich estimator, the bootstraps,
and subsampling. Although we have presented the code for each of the methods
separately  
for instructive purposes, all of these variance estimators can be computed
simultaneously through a single call of the \texttt{comp\_var()} function as
follows:

\begin{lstlisting}[language=R, title={}, caption={}, basicstyle=\footnotesize\ttfamily,
label={nlst:r-all-methods},numbers=none]
# obtain maars obj. w/ all methods
maars_var <- mod_fit %>%
  comp_var(
    boot_emp = list(B = 1e3, m = 505),
    boot_mul = list(B = 1e3),
    boot_sub = list(B = 1e3, m = 200),
    boot_res = list(B = 1e3))
\end{lstlisting}

\nit The \texttt{maars\_var} object (\ie, of class \texttt{maars\_lm}) by
default supports the generic \texttt{print()} and \texttt{summary()} methods in
\texttt{R} that are typically called on \texttt{lm()} objects, for instance. In
fact, for our newly created \texttt{maars\_var} object, a natural first step in
the \texttt{maars} inferential workflow is for the user to simply run the
generic \texttt{print()} method. This can be used to generate a summary of the
OLS estimates and of the assumptions behind each of the variance estimation
methods that have been employed:

\begin{lstlisting}[language=R, title={}, caption={}, basicstyle=\footnotesize\ttfamily,
label={nlst:r-print},numbers=none]
# print obj.
print(maars_var)
\end{lstlisting}

\nit As shown in the left panel of \Cref{fig:print-summary}, this method essentially returns an augmented version of the typical \texttt{print()} called on an \texttt{lm} object. 
Besides the OLS estimates, for each variance estimation methods that has been used in \texttt{comp\_var()}, the assumptions (\eg, homoscedasticity) and parameters chosen (\eg, $n$, $B$, and $m$) are displayed. 
We hope our default feature of explicitly presenting assumptions will become a de facto
standard for most statistical modeling in \texttt{R}.

The computed inferential statistics can be obtained as follows:

\begin{lstlisting}[language=R, title={}, caption={}, basicstyle=\footnotesize\ttfamily,
label={nlst:r-summary},numbers=none]
# summarise obj.
summary(maars_var)
\end{lstlisting}

\begin{figure}[H]
    \begin{subfigure}{.48\textwidth}
    \centering
    \includegraphics[width=\textwidth]{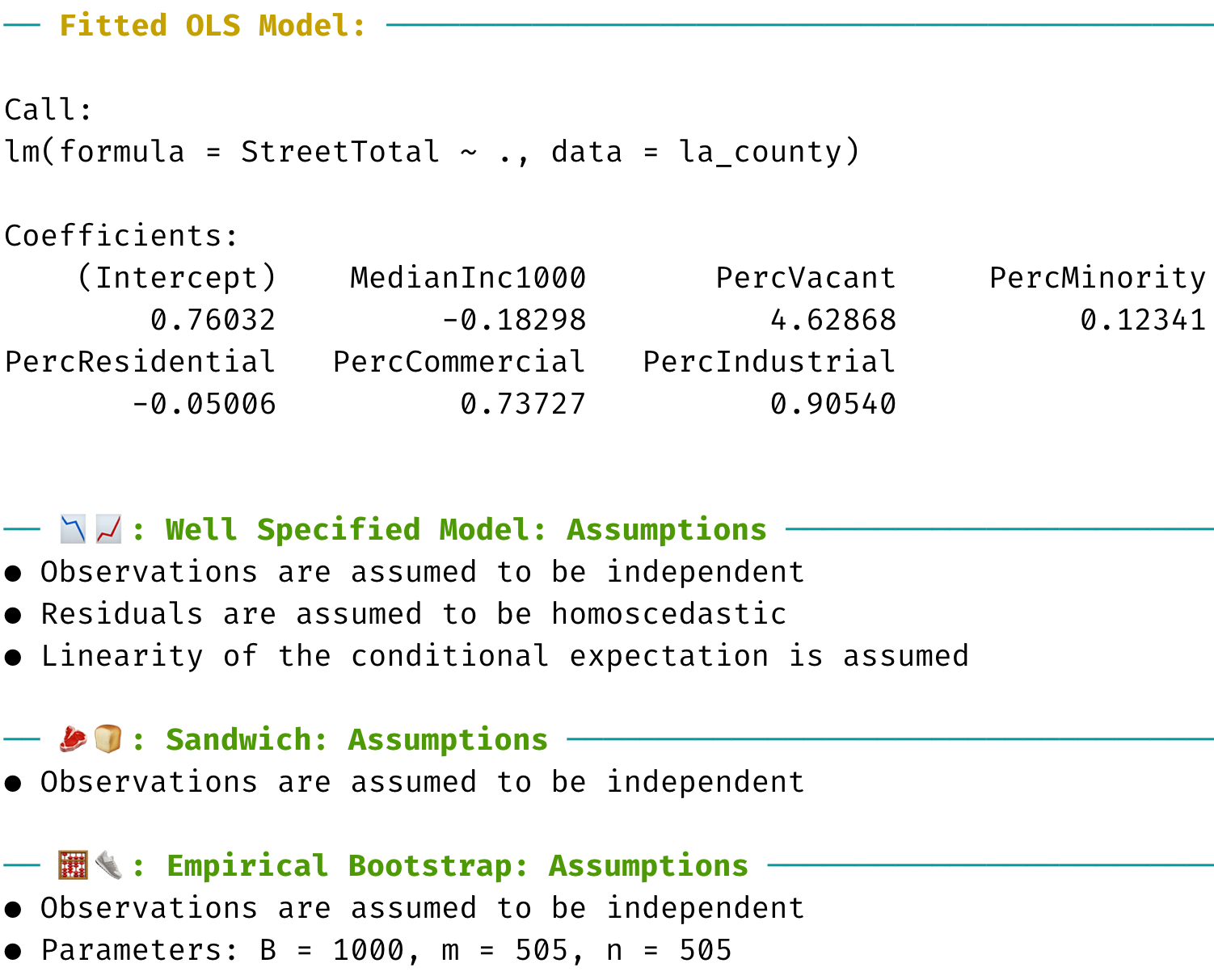}
    \label{fig:print-maars-var-output-split}
    \end{subfigure}
    \hfill
    \begin{subfigure}{.48\textwidth}
    \centering
    \includegraphics[width=\textwidth]{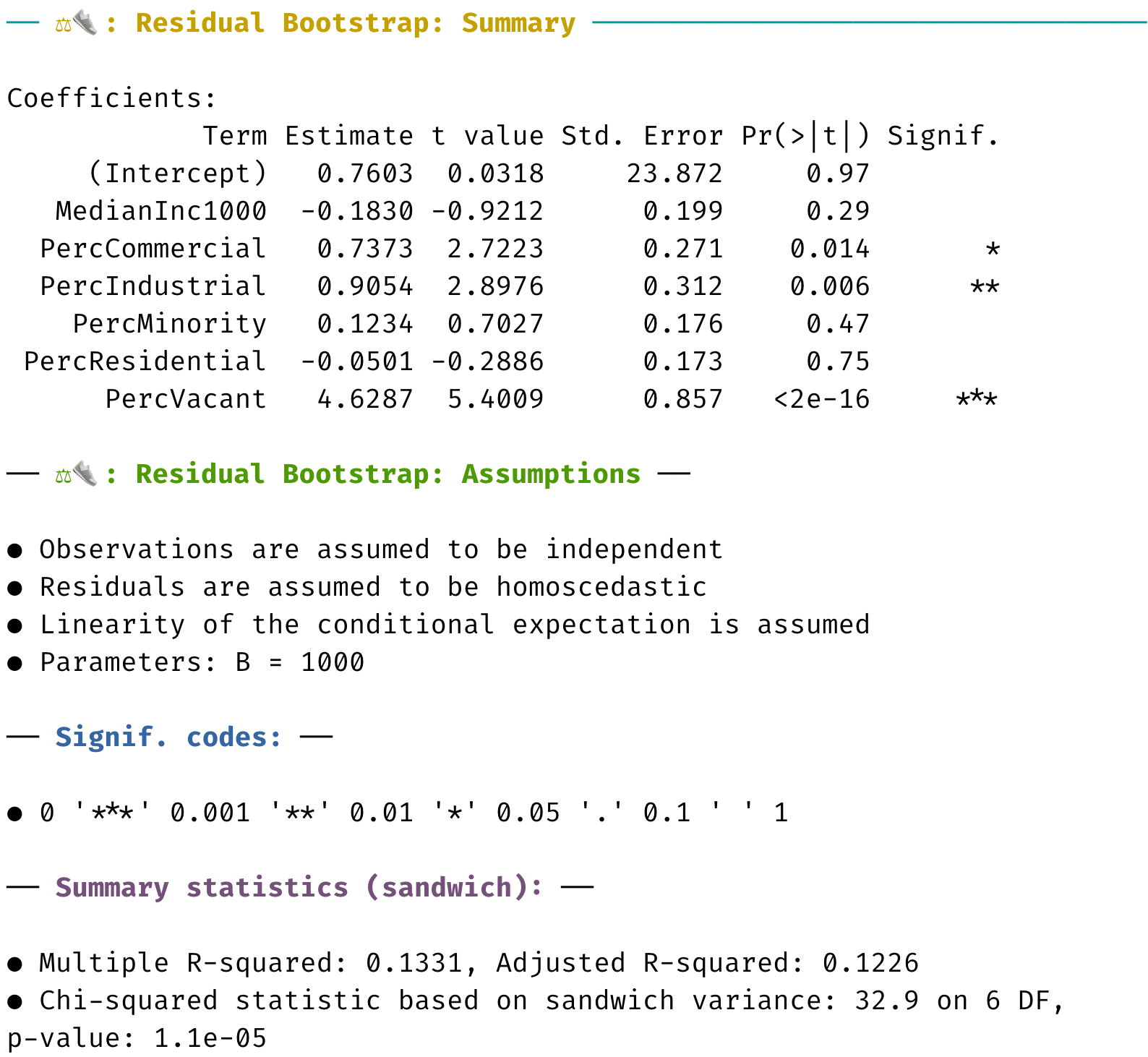}
    \label{fig:summary-output-split}
    \end{subfigure}
    \caption{Model diagnostics for the \texttt{maars\_var} object. Left panel: Partial console output of \texttt{print(maars\_var)}. Right panel: Partial console output of \texttt{summary(maars\_var)}.}
    \label{fig:print-summary}
\end{figure}

As shown in the right panel of \Cref{fig:print-summary}, calling the \texttt{summary()} method 
on a \texttt{maars\_lm} object prints a stacked series of tibbles in a similar format as 
returned by the \texttt{tidy()} function in the \texttt{broom} package \citep{broom2020cran}.
They are based on the different variance estimation methods, together with the related sets 
of assumptions.
The result of a $\rchi^{2}$ test for testing the null hypothesis that the OLS coefficients are 
all equal to 0 is reported at the bottom of the printed output. We make use
of the \texttt{cli} package \citep{csardi2021clicran} to produce well-structured 
console output for both the \texttt{print()} and \texttt{summary()} methods.

In order to simplify downstream manipulation of the results, 
the generic methods present in \texttt{maars} have \emph{tidy analogues}. For example, the user 
can obtain the set of OLS estimates and variance estimates in 
the form of a tibble by running \texttt{get\_summary(maars\_var)}.
More specifically, these functions begin with a consistent \texttt{get\_} prefix
and return tidy \texttt{tibble} objects for data, \texttt{ggplot2} objects for plots.
By tidy tibble output we mean data in which every column is a variable, every row 
is an observation, and every cell is a single value \citep{wickham2014tidydatajoss}.
We believe that this tidy grammar for function naming
enables efficient further analysis of results
using the \texttt{tidyverse} or similar \texttt{R} packages.

\section{Visual diagnostic tools}\label{sec:diagnostic-tools}

\begin{figure}[H]
\begin{subfigure}{.5\textwidth}
\centering
\includegraphics[width=\textwidth]{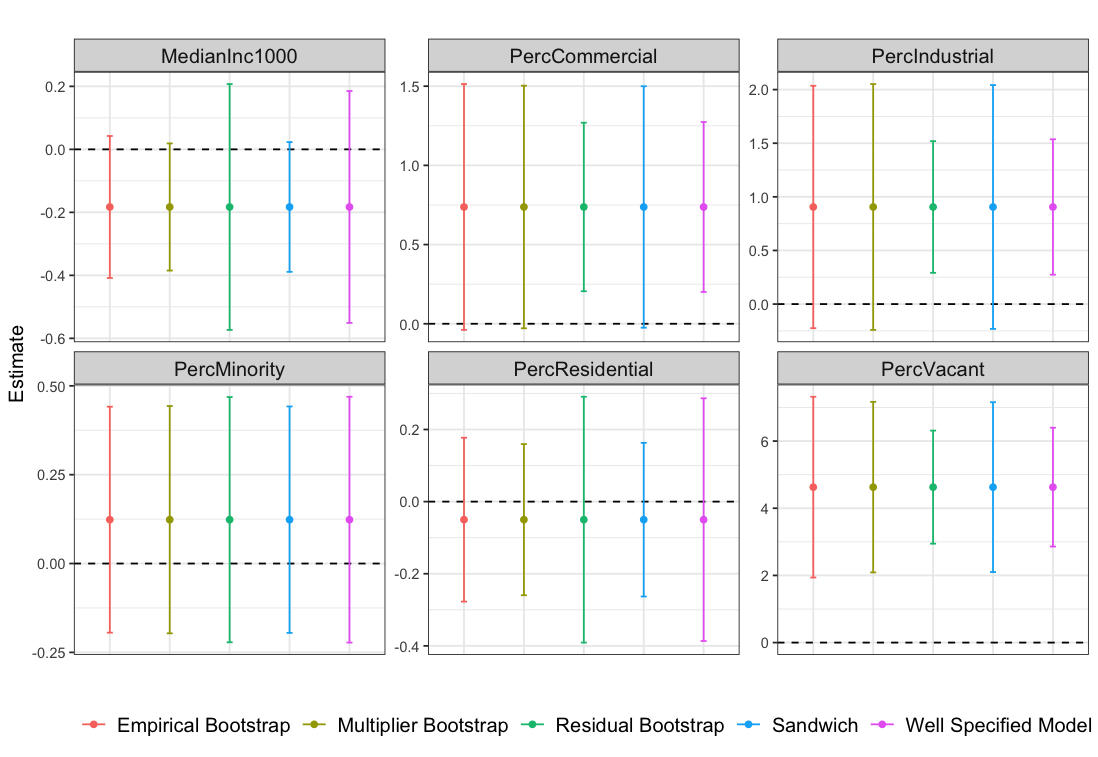}
\label{fig:conf_int_comparison}
\end{subfigure}
\begin{subfigure}{.5\textwidth}
\centering
\includegraphics[width=\textwidth]{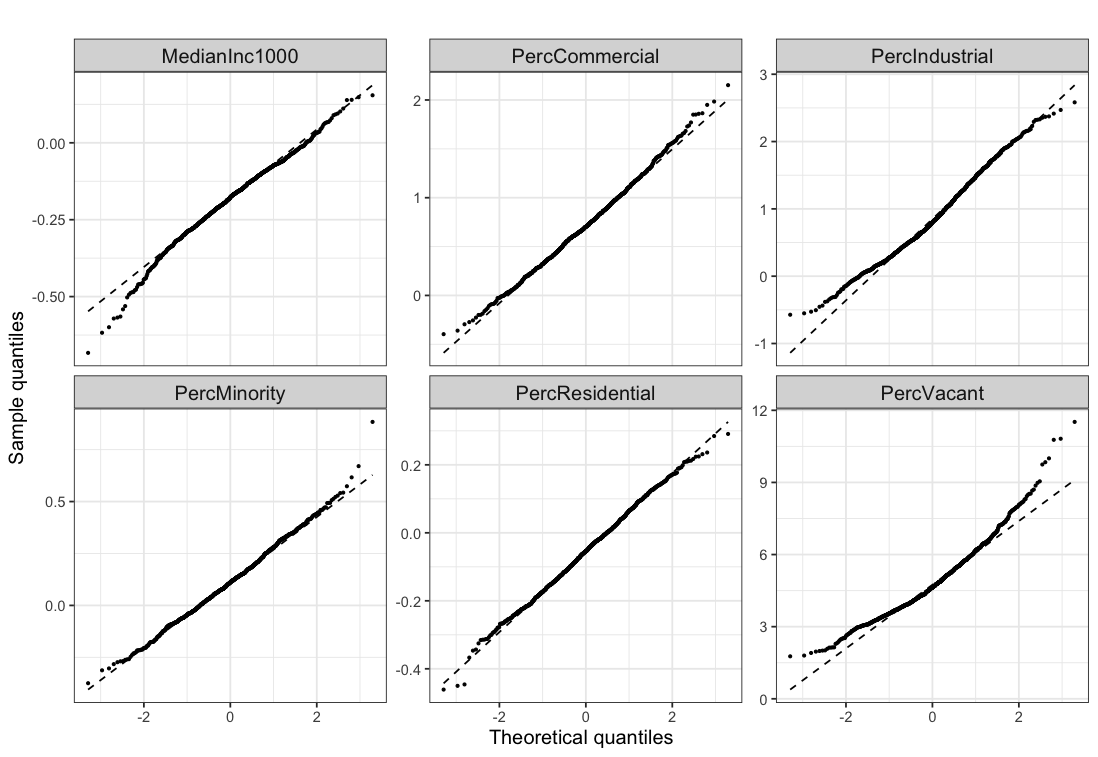}
\label{fig:r_qqnorm_emp}
\end{subfigure}
\caption{Model diagnostics for the \texttt{maars\_var} object. Left panel: 95\% confidence intervals based on the different variance estimation methods. Right panel: Normal Q-Q plot based on the OLS estimates obtained via empirical bootstrap.}
\label{fig:model-diagnostic-plot}
\end{figure}

Using the \texttt{maars\_var} object that we created in \Cref{sec:variance-estimation-maars}, the user can readily inspect the model's fit via model diagnostic plots by calling the \texttt{plot()} method and its related tidy analogue \texttt{get\_plot()}.
These functions return eight graphical diagnostics. 
Six of them correspond to the same output that is generated by calling \texttt{plot(mod\_fit)}. These are the same plots returned from the classical \texttt{lm()} object but returned as 
\texttt{ggplot2} objects. 
The two additional diagnostics are presented in \Cref{fig:model-diagnostic-plot}.
The diagnostic in the left panel displays the widths of the different 95\% confidence intervals for the OLS coefficients estimates. This graphical tool allows the user to visually compare the variance estimators computed using \texttt{comp\_var}. Recall that the \texttt{lm()} and the residual bootstrap based variance estimator are only valid under a well-specified linear model, whereas the empirical, multiplier bootstrap, and subsampling variance estimator are model-free. Therefore, the visual comparison of estimated variances (as in the left panel of Figure~\eqref{fig:model-diagnostic-plot}) already provides evidence of misspecification, if any.
The last diagnostic, which is shown in the right panel of Figure~\ref{fig:model-diagnostic-plot}, presents the distribution of the OLS estimates on the bootstrapped data sets. 
Through this tool, the user can compare the distribution of the estimates via a normal Q-Q plot to visually check whether sample size is large enough to justify the asymptotic regime. 

The \texttt{maars} package also includes the three graphical diagnostic tools proposed in \cite{buja2019modelsasapproximationspart2}. 
When the regression model is correctly specified, the regression coefficients do not depend on the
distribution of $X$; see \citet[Proposition~4]{buja2019modelsasapproximationspart2}.
Equivalently, variations in the OLS estimates under reweighting of $X$ would suggest that the model is not well-specified. 
In \texttt{maars}, we have implemented 
the graphical model diagnostics tools through which one
can investigate whether the OLS assumptions are broken. 
These tools, which were proposed by~\cite{buja2019modelsasapproximationspart2}, supplement of the model diagnostics returned by \texttt{lm()}.

The key idea behind these model diagnostics is to study how the OLS estimates vary 
under arbitrary reweighting of $X$. 
It is hard visualize such a diagnostic with arbitrary reweighting and hence we restrict to reweighting along certain regressors, as in~\cite{buja2019modelsasapproximationspart2}. Although restrictive, 
the resulting plots provide intuitive understanding of misspecification in data.
For reweighting along the $j$-th regressor $X(j)$, $1\le j\le d$, we do the following two operations:
\benum[label=\textbf{DT.\arabic*}, align=left, start=1]
    \item \label{itm:dt1} Select a grid of $K$ values (or centers) $c_{1,j},\dots,c_{K,j}$ on the support of $X(j)$.
    \item \label{itm:dt2} For each center $c_{\ell, j}, 1\le \ell\le K$, compute the weighted least squares estimator
    \begin{align}
        \widehat{\beta}_{\ell} = \argmin_{\theta\in\mathbb{R}^d}\sum_{i=1}^n (Y_i - X_i^{\top}\theta)^2e^{-{(X_{i}(j) - c_{\ell,j})^2}/{(2\gamma^2)}}.
    \end{align}
    Intuitively, this is the least squares estimator that gives more weight to 
    observations that have $j$-th regressor values close to $c_{\ell,j}$.
\eenum
The procedure consists of fitting $K$ regressions. This will give us one curve and for comparison, we also run the same procedure on bootstrapped data leading to $\widehat{\beta}^*_{\ell}, 1\le \ell\le K$.
The tibble containing the OLS estimates based on the reweighting procedure can be obtained using the following function:

\begin{lstlisting}[language=R, title={}, caption={}, basicstyle=\footnotesize\ttfamily,
label={nlst:r01-mod-diagnostics},numbers=none]
# regression based on reweighting of regressors
coef_rwgt <- mod_fit %>% 
	diag_fit_reg_rwgt(B = 300, terms_to_rwgt = names(la_county)[-1])
\end{lstlisting}

In the package, we have developed two types of grids for \ref{itm:dt1}.
The default grid of reweighting centers in \ref{itm:dt1} is based on the deciles of $X(j)$ as in~\cite{buja2019modelsasapproximationspart2}.
Alternatively, the user can select a grid of evenly spaced values between $X_{(1)}(j)$ and $X_{(n)}(j)$.
For \ref{itm:dt2}, \texttt{maars} uses
$\gamma = (\sum_{i=1}^n (X_i(j) - \overline{X}(j))^2/(n-1))^{1/2}$. 
To assist the data scientist in interpreting the 
results of the refitting procedure, 
\cite{buja2019modelsasapproximationspart2} have proposed using the following three graphical model diagnostics tools.
\begin{itemize}[leftmargin=*]
    \item {\bf focal slope}: Consider changes in coefficient of interest $\widehat{\beta}_{\ell}(k)$ to the
    reweighting of each of the regressors $X(j)$, $1\le j\le d$; this provides insights into the interactions between regressor $X(k)$ and all other regressors $X(j)$. 
    \item {\bf nonlinearity detection}: Consider changes in $\widehat{\beta}_{\ell}(k)$ to the reweighting of its own regressor $X(k)$; this provides insights into marginal nonlinear behaviors of the response surface. 
    \item {\bf focal reweighting variable}: Consider changes in all coefficients $\widehat{\beta}_{\ell}(j)$ to the reweighting of a given regressor $X(k)$.
\end{itemize}
For further information and interpretations of these types of diagnostics, see 
\citet[Section~5]{buja2019modelsasapproximationspart2}. 
The three types of graphical diagnostics can be obtained in \texttt{maars} as follows.

\begin{lstlisting}[language=R, title={}, caption={}, basicstyle=\footnotesize\ttfamily,
label={nlst:r02-mod-diagnostics},numbers=none]
# focal slope for "PercVacant"
mod_fit %>% diag_foc_slope(coef_rwgt,'PercVacant')
# nonlinearity detection
mod_fit %>% diag_nl_detect(coef_rwgt)
# focal reweighting var. for "PercVacant"
mod_fit %>% diag_foc_rwgt(coef_rwgt,'PercVacant')
\end{lstlisting}

\begin{figure}[H]
\centering
  \includegraphics[width=0.7\linewidth]{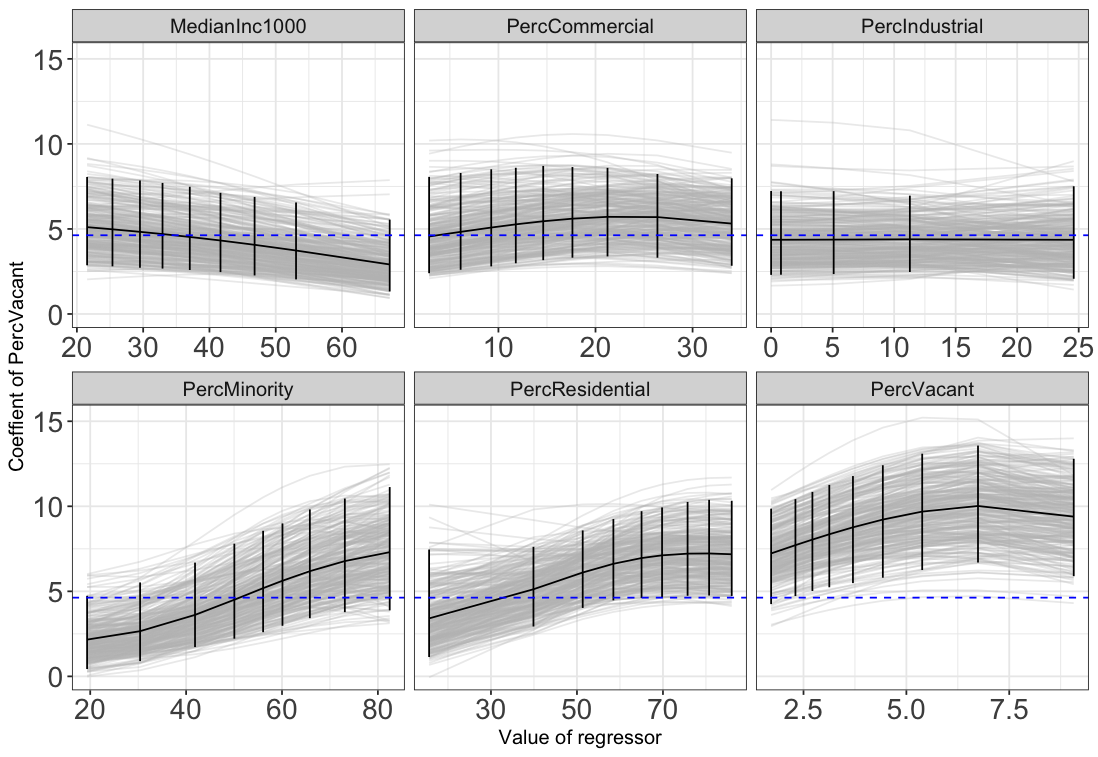}
    \caption{Focal slope: the $x$-axis represents the grid mentioned in \ref{itm:dt1}. Intercept is excluded.}
    \label{fig:r-focal-slope}
\end{figure}

A sample output of the focal slope diagnostics is shown in~\Cref{fig:r-focal-slope}. 
These plots show how the OLS estimate of the prevalence of vacant lots ({\it PercVacant}) in our running example for 300 bootstrapped data sets (gray lines) vary under reweighting of all regressors (plot panel titles). 
The changes observed in the means of the estimates (black lines in the middle of each panel) indicate that the model is unlikely to be well-specified.
In particular, these plots suggest the presence of an interaction between {\it PercVacant} and other regressors such as {\it PercMinority}, {\it PercResidential}, and {\it MedianInc (\$1000)}.
In the plot in the bottom right corner, which depicts the estimate of {\it PercVacant} under reweighting of its own regressor, we observe that the reweighted estimates are far larger than the unweighted ones on the original data (blue dashed line).

\section{Teaching by example: vignettes}\label{sec:teaching-by-example-vignettes-grammar}

As previously noted, \texttt{maars} is developed as a pedagogical (and research)
tool to convey the rich techniques and concepts for OLS under model
misspecification. Specifically, per~\ref{itm:dp4} we emphasize teaching these
concepts through guided case studies using tidy \texttt{maars} workflows. Our goal is
to showcase these case studies as vignettes on our official package website. 

These vignettes take two main forms. First, we take the core \textit{research}
papers upon which \texttt{maars} is based
and reproduce tables and plots from them \citep{buja2019modelsasapproximationspart1, buja2019modelsasapproximationspart2}. The intent here is to make the latest
theoretical research in OLS under model misspecification accessible to the data
scientist in a hands-on manner. Second, we provide \textit{lesson plan}
vignettes for these inferential techniques. For example, we know that both
empirical and multiplier bootstrap are valid under model
misspecification, but how different are they in practice? A systematic way to
approach this question in a classroom setting (and beyond) is to observe that
they contain the number of bootstrap replications, $B$, as a common parameter.
Under a simulated misspecified linear model, we demonstrate using a tidy
\texttt{maars} workflow how to plot the confidence interval coverage, and average
confidence width for these two estimators, as a function of the common parameter
$B$. The tidy \texttt{maars} grammar helps illustrate how to efficiently and
consistently prototype these concepts in \texttt{R}, particularly to new
students. We currently have two such case studies available and plan to
extend this to a much a larger repository of research and pedagogical vignettes
in the near future.

\section{Open-source best practices}\label{sec:open-source-best-practices}

Our goal in developing the \texttt{maars} package is for it to be a standard
\texttt{R} workflow for OLS inference under the model misspecification framework
described in \cite{buja2019modelsasapproximationspart1,
buja2019modelsasapproximationspart2}, and related literature. Per~\ref{itm:dp5},
we have strived to implement and adapt to open-source best practices to ensure 
an inclusive community for all users and contributors.

In developing the \texttt{maars} package, we have setup benchmarking
using the~\texttt{bench} package~\citep{bench2020cran}. We plan to conduct 
high-precision benchmarking of the core \texttt{maars} 
package functionality, and to assist in future code efficiency profiling efforts.
Similarly, we currently test our statistical functionality and error handling
using the \texttt{testthat} package~\citep{testthat2011}. Moreover, testing 
is built into the development cycle through cross-platform continuous integration 
using Github Actions. 
We use the \texttt{tidyverse} contributing guide and code of 
conduct to ensure that new contributors 
have ample guidance on the \texttt{maars} package
development practices.

\section{Conclusion and future work}\label{sec:conclusion-and-future-work}

In this paper, we introduce the \texttt{maars} package functionality to perform OLS
inference under model misspecification. The \texttt{maars} package is designed
to implement the key inferential ideas from
\citet{buja2019modelsasapproximationspart1, buja2019modelsasapproximationspart2}
in a tidy \texttt{R} workflow. 

The \texttt{maars} package is built on five core software design principles (see \ref{itm:dp1}--\ref{itm:dp5}). 
These design principles affect the day-to-day
model-free inferential workflow. They also influence the way
in which users can contribute to the existing \texttt{maars} functionality. The
main tools for statistical inference in this framework are based on
different variance estimation methods. These variance estimation tools in
\texttt{maars} include closed-form variance estimators and standard errors
computed using resampling techniques. By default \texttt{maars} provides valid
hypothesis testing under model misspecification, \ie, reporting $\rchi^{2}$ tests
rather than $F$ tests. To help diagnose issues under the misspecified setting,
there are a number of visualization tools readily accessible to
\texttt{maars} user. These currently include the focal slope graphs of
regressor variables \citep{buja2019modelsasapproximationspart2} and the Q-Q plots.

The \texttt{maars} package is still in active development and the final version
we envision includes more tools related to misspecification. 
Implementation of analysis of variance (ANOVA) under misspecification and conformal inference based 
prediction~\citep[Section~2.3]{vovk2005algorithmiclearningrandworld} 
is in testing phase and will be added to the package soon. Further future
work includes other variants of inference under dependent and time series data (beyond subsampling) 
as well as cluster robust standard errors.
Finally,
we want to extend the functions to GLMs, the Cox proportional hazards model, and
inference after data exploration, \eg, PoSI~\citep{kuchibhotla2020}.
We intend to demonstrate new
package functionalities through more illustrative case study vignettes. 

\section*{Acknowledgements}\label{sec:acknowledgments}
The authors would like to thank the members
of the Larry's group at the Wharton Statistics Department
for their valuable feedback and encouragement during this work 
(listed alphabetically by surname): 
Richard Berk, Andreas Buja, Junhui Cai, Edward George,
Elizabeth Ogburn, and Linda Zhao. 
The authors would also like to thank Alex Reinhart from 
the CMU Department of Statistics \& Data Science for 
sharing valuable insights which informed our package development.
\bibliographystyle{plainnat}
\bibliography{refs}

\end{document}